\documentclass[epj]{svjour}

\usepackage{graphics}
\usepackage{amsmath}
\usepackage{amssymb}
\begin{document}
\title{Segregation in desiccated sessile drops
of biological fluids}
\author{Yu.~Yu.~Tarasevich \and D.~M.~Pravoslavnova}% etc
\institute{Astrakhan State University, 20a Tatishchev St.,
Astrakhan, 414056, Russia}
\date{Received: date / Revised version: date}
% The correct dates will be entered by Springer
%
\abstract{ It is shown here that concurrence between advection and
diffusion in a drying sessile drop of a biological fluid can produce
spatial redistribution of albumen and salt. The result gives an
explanation for the patterns observed in the dried drops of the
biological fluids.
 % end of PACS codes
} %end of abstract
\maketitle
\section{Introduction}\label{intro}

\begin{figure}
\resizebox{\columnwidth}{!}{%
\includegraphics{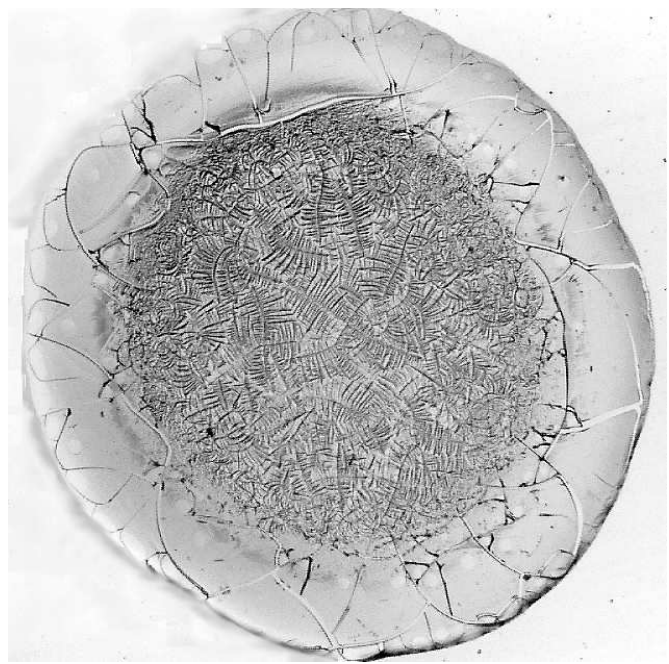}
} \caption{A dried drop of the multi-component fluid consisting of
NaCl (0.9~\%), albumen (9~\%), and water. This composition and the
concentrations are typical for the blood serum of the healthy
persons.\label{fig:sample}}
\end{figure}

The dried droplets of the biological fluid have very complex
structure. In particularly, there are protein ring at the periphery
and dendritic crystals in the central area of the sample
(Fig.~\ref{fig:sample}). The visual appearance of a sample is used
for diagnosing a wide range of diseases. Rather clear relationships
of the 'pattern~--- pathological processes' type have been revealed
\cite{book2}.

The priority in the investigation of the dehydration
self-organization phenomenon in biological fluids belongs to
Rapis~\cite{book1}. While seeming simple, the effect turned out to
be extremely complicated and to involve a number of interrelated
processes of different nature.

The biological fluids are complex colloidal systems. Even the
slightest variation in the composition of a liquid leads to total
changes in the dynamics of phase transitions during drop
drying~\cite{Yakhno}. Despite the application of the effect to
practical medical diagnostics \cite{book2}, and the drastic
improvement of our understanding of pattern formation, complex
non-linear dynamics of pattern formation in drops of biological
fluids is mostly unclear. Nevertheless, the performed analysis
allows to conclude that the main effects observed in the dehydration
of biological fluids are typical for colloidal solutions in general
and can be described in the framework of conventional
approaches~\cite{tarasevich1}.

This work is an part of the research on mathematical modeling of
pattern formation in a biological fluid upon desiccation, with a
special emphasis to evaporation, capillary flow \cite{tarasevich2},
effect of diffusion \cite{tarasevich3,tarasevich2007} and sol-gel
phase transition. The main scientific goal of the research is
determination the relationship between the pathological processes
occurring in an organism, the variation of the physical and
physicochemical properties of the biological fluids caused by these
processes, and the kind of the patterns produced in the drying of
the sessile droplet of a biological fluid. The results presented in
this paper allow to give the fundamental explanation of spatial
component redistribution as a consequence of diffusion and capillary
flow.

\section{Model}\label{sec:model}
\subsection{Theory of solute transfer}\label{subsec:transfer}

Detailed investigation of of solute transfer in the desiccated drops
was performed by Deegan et al.~\cite{Deegan}. In their theory, an
outward flow in a drying drop of liquid is produced when the contact
line is pinned so that liquid that is removed by evaporation from
the edge of the drop must be replenished by a flow of liquid from
the interior. This flow is capable of transferring 100\% of the
solute to the contact line and thus accounts for the strong
perimeter concentration of many stains. Furthermore, the theory
relies only on a generic property of the substrate-solvent
interaction, the presence of surface roughness or chemical
heterogeneities that produce contact line pinning, and therefore it
accounts for the ubiquitous occurrence of ringlike stains.

Indeed, as soon as evaporation begins, particles deposit onto the
substrate, resulting in a strong anchoring of the three-phase line.
In the case of biological fluid, proteins are adsorbed on the
substrate, leading to a strong anchoring of the triple line.

Recently, segregation in multi-component ceramic colloids during
drying of droplets was investigated \cite{Wang2006}.

If the size of the solute particles is small, diffusive currents
become comparable to the advective currents \cite{Deegan}.

Biological fluid widely used in the medical tests is blood serum. It
consists of the proteins (mostly albumen)~--- about 9~\%, NaCl~---
0.9~\%, and water. It is shown in this work, that concurrence
between diffusive and capillary flows can produce more uniform
distribution of salt along a drop diameter, but has not essential
effect on albumen distribution.

\begin{figure}
\resizebox{\columnwidth}{!}{%
\includegraphics{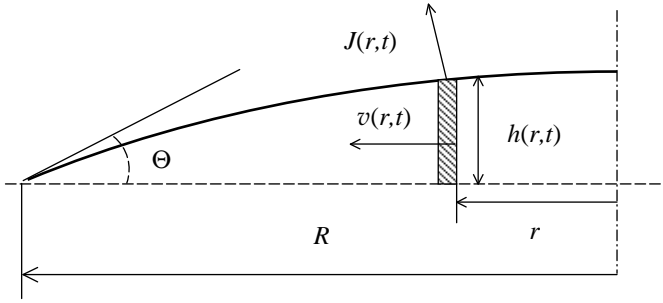}}
\caption{Schematic of relevant parameters for the
theory.\label{fig:cons}}
\end{figure}

The conservation of fluid determines the relationship \cite{Deegan}
between the vertically averaged radial flow of the fluid, $v$, the
position of the air-liquid interface, $h$, and the rate of mass loss
per unit surface area per unit time from the drop by evaporation,
$J$:
\begin{equation}
\label{eq:cons} \frac{\partial }{\partial t}\left( {\rho h} \right)
+ \frac{1}{r}\frac{\partial }{\partial r}\left( {r\rho hv} \right) +
J\sqrt{1+\left( \frac{\partial  h}{\partial  r}\right)^2} = 0,
\end{equation}
where $\rho $ is the density of the liquid. All quantities are
supposed to be dependent both of distance $r$, and of time $t$. For
additional explanation see Fig.~\ref{fig:cons}.

The flow velocity can be expressed from \eqref{eq:cons} by rewriting
it in the integral form~\cite{Deegan}
\begin{equation}
\label{eq:v} v = - \frac{1}{\rho r h} \int_0^r dr \left( J \sqrt{1 +
\left( \frac{\partial h}{\partial r}\right)^2} + \rho \frac{\partial
h}{\partial t} \right) r .
\end{equation}

The conservation of solute gives
\begin{equation}
\label{eq2} \frac{\partial }{\partial t}\left( {\rho ch} \right) +
\frac{1}{r}\frac{\partial }{\partial r}\left( {r\rho hc\tilde {v}}
\right) = 0,
\end{equation}
where $c$ is concentration, $\mathbf{\tilde {v}} = \mathbf{v} +
\mathbf{v_{\text{d}}}$ is velocity including both advective term,
$\mathbf{v}$, and diffusive one,
$\mathbf{v_{\text{d}}}$,~\cite{landau}. Here $c{\rm {\bf
v}}_{\text{d}} = - D \mathrm{grad}\, c$, where $D$ is the diffusion
constant for solute in solvent. Then
\begin{equation}
\label{eq3} v_{\text{d}} = - \frac{D}{c}\frac{\partial c}{\partial
r}.
\end{equation}

Master equation describing spatial and temporal solute dynamics is
\begin{multline}
\label{eq:diff} \frac{\partial c}{\partial t} =
\frac{1}{r}\frac{\partial }{\partial r}\left( {r D \frac{\partial
c}{\partial r}} \right) + \\ \frac{\partial c}{\partial r}\left(
\frac{D}{ h} \frac{\partial  h}{\partial r}  + \frac{D}{\rho }
\frac{\partial \rho}{\partial c}\frac{\partial c}{\partial r}- v
\right) + \frac{cJ}{\rho h}\sqrt{1+\left(\frac{\partial h}{\partial
r}\right)^2}.
\end{multline}

One needs to know the evaporation law, the drop profile, and the
vertically averaged radial flow of the fluid inside the drop to
solve the master equation.

\subsection{Simplifying assumptions}\label{subsec:assumptions}

We will suppose that
\begin{itemize}
     \item air--liquid interface has a spherical cap shape induced by surface
     tension~\cite{Deegan};
     \item drop apex height steadily decreases with constant velocity $v_0$:
     $h(0,t) = h_0 - v_0 t$, where $h_0$ is initial height of drop apex~\cite{Pauchard}; then
$J = \rho v_0 /2$;
     \item in a multicomponent fluid the components do not influence
     on each other;
     \item diffusion coefficient for each component $D$ is constant;
     \item variation of the solution density $\rho$ \emph{vs.}
     concentration $c$ is rather small, so we will suppose constant
     density everywhere, but the term containing ${\partial \rho}/{\partial c}$,
     because this term may be very large even for small variations of the density.
\end{itemize}

Detailed description of the main assumption can be found in
Ref.~\cite{Popov}.

\subsection{Evaporation models}\label{subsec:evapor} We could not find
any experimental data about spatial distribution of vapor flux near
the drop surface, the theoretical models are rather controversial
and contradictory.

The analogy between diffusive concentration fields and electrostatic
potential fields (they both satisfy Laplace's equation) is widely
used for calculation of the vapor flux of a sessile
droplet~\cite{Deegan,Hu2002,Mollaret2004}. In particularly, this
approach was used to describe the vapor flux of very thin
evaporating droplets~\cite{Popov,Poulard}. In this model, the
evaporation flux is almost uniform all over free surface except very
narrow region near the drop edge~\cite{Hu2002}. Nevertheless, there
is a singularity at the contact line. The divergence at the edge is
nonphysical and can be removed by using a smoothing
function~\cite{Poulard}.

An alternative method for determining the evaporative mass flux is
to use heat transfer analysis~\cite{Anderson1995}. Therefore, this
form cannot be applied to a droplet in which the contact line is
pinned because the presence of colloidal particles affects  the
evaporation near the edge~\cite{Fischer2002}. To mimic this effect
the evaporation function was modified~\cite{Fischer2002}.

On the other hand, there is a region near the contact line with high
evaporation rate~\cite{Takhistov}.

Following Ref.~\cite{Parisse96} the simplest possible kind of vapor
flux is assumed in the present work, i.e. a constant evaporation
rate all over free surface.

\subsection{Dimensionless form of the master
equation}\label{subsec:dimless}

Using above assumptions, one can rewrite master equation
\eqref{eq:diff} in the dimensionless form
\begin{multline}
\label{eq:masterdl} \frac{\partial c}{\partial \tau } =
\frac{\mathbb{D}}{x}\frac{\partial }{\partial x}\left(
{x\frac{\partial c}{\partial x}} \right) +\\ \frac{\partial
c}{\partial x}\left( \frac{\mathbb{D}}{L}\frac{\partial L}{\partial
x} + \frac{\mathbb{D}}{\rho } \frac{\partial \rho}{\partial
c}\frac{\partial c}{\partial x} - kv\right)  + \frac{kc}{2L}\sqrt {1
+ \left( {\frac{\partial L}{\partial x}} \right)^2},
\end{multline}
where $\mathbb{D}= h_0 D/(v_0 R^2)$ is dimensionless parameter,
$\tau = t v_0/h_0$ is dimensionless  time, $x = r/R$ is
dimensionless distance, and $k = h_0/R$. $L( {x,\tau })$ is
dimensionless profile of the drop written as

$$L\left( {x,\tau } \right) = \sqrt {\left( {\frac{A^2 + 1}{2A}}
\right)^2 - x^2} - \frac{1 - A^2}{2A},$$ where $A(\tau ) = k(1 -
\tau )$ and $$B(x,\tau ) = \sqrt {\left( {A^2(\tau ) + 1} \right)^2
- \left( {2A(\tau )x} \right)^2} - A^2(\tau ) - 1.$$

The dimensionless velocity can be written as
\begin{equation*}
v(x,\tau ) = \frac{\left( {A^2(\tau ) + 1} \right)\left(
{\frac{B(x,\tau )}{2A^2(\tau )} + x^2} \right)}{4A(\tau )xL(x,\tau
)}.
\end{equation*}

\section{Results and discussion}\label{sec:results}

In medical diagnostics, drops of a biological liquid in volume
10--20~$\mu $l are used. The drops are placed onto the strictly
horizontal glass. Thus diameter of a droplet is 5--7~mm and $h_0
\approx 1$~mm~\cite{book2}. Hence, $k=0.33$. Diffusion coefficients
are $D_\text{a} = 7.7 \cdot 10^{-11}$~m$^2$/sec (albumen) and
$D_\text{s} = 1.5 \cdot 10^{-9}$~m$^2$/sec (NaCl). Experiments on
measurement of height of a drying up drop show, that $v_0 \approx 4
\cdot 10^{ - 7}$~m/sec~\cite{Annarelli2001}. Then dimensionless
parameters are $\mathbb{D}_\text{a} \approx 0.03$ (albumen) and
$\mathbb{D}_\text{s} \approx 0.6$ (NaCl). We suppose that salt has
no effect on solution density because of very small concentration.
Then density of solution is linear function of albumen concentration
$\rho(c) = 1000 + 300 c$~kg/m$^3$ and $\frac{1}{\rho }
\frac{\partial \rho}{\partial c} = 0.3$ \cite{Chick}.

\begin{figure*}
\resizebox{\textwidth}{!}{%
\includegraphics{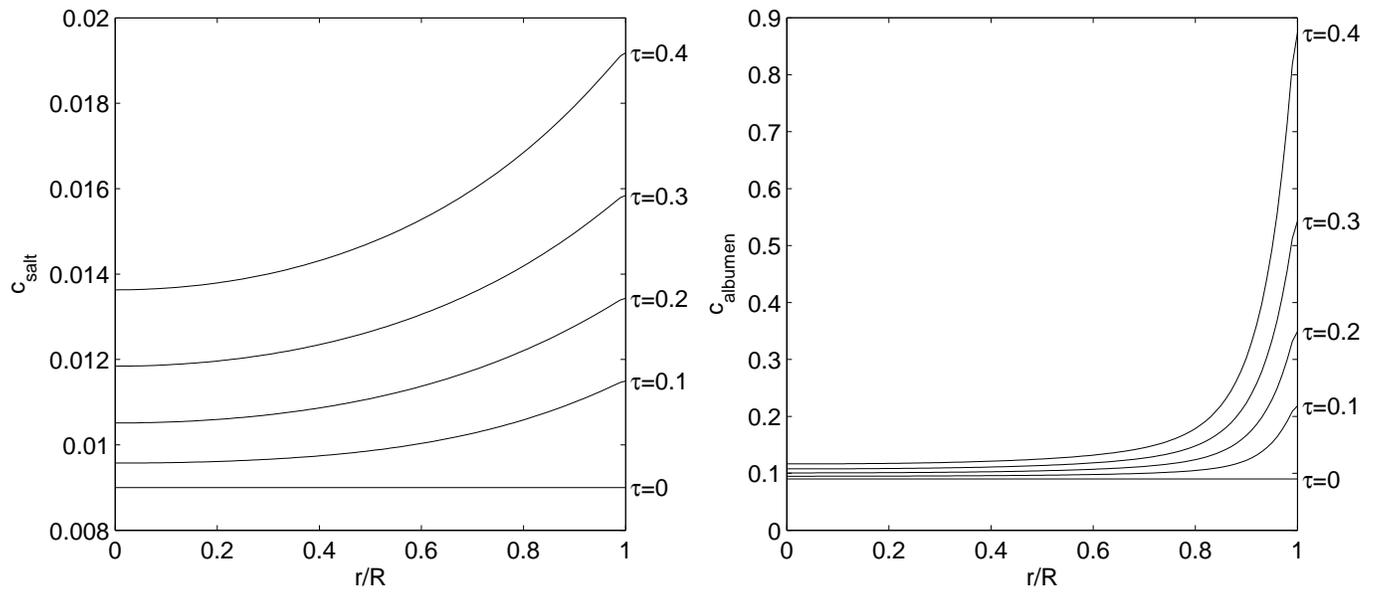}}
\caption{Dynamics of the spatial distribution of the components
inside a desiccated drop of biological fluid.\label{fig:ratio}}
\end{figure*}

Equation \eqref{eq:masterdl} was solved numerically. Presented in
Fig.~\ref{fig:ratio} results show that diffusive processes can
prevent carrying out of salt on edge by capillary current. At the
same time, diffusion has not essential influence on spatial
distribution of albumen. As a result the edge of a sample consists
mainly from albumen, while its central part is composed both from
salt and from albumen. This result confirms qualitative conclusions
obtained earlier using simplest
model~\cite{tarasevich3,tarasevich2007}. One can see that near the
drops edge there is rather narrow region with high concentration of
albumen. Phase transition of albumen from sol to gel have to occur
in this region.

\section{Acknowledgments}
The authors are grateful to the Russian Foundation for Basic
Research for funding this work under Grant No. 06-02-16027-a.

\end{document}